# Current fluctuations of polymeric chains

Kamil Walczak [1]

Institute of Physics, Adam Mickiewicz University
Umultowska 85, 61-614 Poznań, Poland

Coherent electron transport is investigated in a molecular device made of polymeric chain sandwiched between two metallic electrodes. Molecular system is described by a simple Hückel model, while the coupling to the electrodes is treated through the use of Newns-Anderson chemisorption theory. Transport characteristics and noise power are calculated in two response regimes: linear and nonlinear, respectively. Here is shown a strong dependence of the shot noise on: (i) the length of the polymeric chain and (ii) the strength of the molecule-to-electrodes coupling. In particular, detailed discussion of Poissonian to sub-Poissonian crossover in the noise spectra is included. Presented algorithm allows to calculate the lowest possible level of current fluctuations (due to Pauli exclusion principle) in designing molecular devices.

**I. Introduction**

Miniaturization trend of microelectronic circuit components, stimulated by advances in the computer industry, will soon reach the scale of individual molecules (or even atoms). Future devices could be composed of two metallic electrodes (source and drain) joined by a molecular wire (bridge). Under the influence of applied bias, the current can flow through the electrode/molecule/electrode system. Indeed, in the past decade electronic transport through a variety individual molecules has been studied as well theoretically as experimentally. The full knowledge of the electrical conduction in such devices is not completed yet, but their transport properties are affected by: (i) the internal structure of the molecule, (ii) the nature of its coupling to the electrodes and (iii) the location of the Fermi level of the metallic electrodes relative to the energy levels of the molecule.

Existing calculations of transport phenomena in nanodevices have mostly been focused on current-voltage (I-V) characteristics only. However, current fluctuations (of thermal or quantum origin) also play an important role in the nanometer scale, especially as the main factor that blurs the signal detection. In order to design molecular devices effectively we have to have an efficient algorithm to calculate their noise power. According to the state of our knowledge, there exists one simple approach to that problem, which allows us to calculate the lowest possible noise level in nanodevices [1]. So-called shot noise (the steady-state current fluctuations) is a direct consequence of the charge quantization and generally is unavoidable even at zero temperature limit (as the only source of electrical noise). Moreover, the noise power is a new transport property, since it contains additional information (which is not included in the conductance spectra) about electron correlations [2]. It also should be mentioned, that conductance fluctuations produced by diagonal and off-diagonal disorders in molecular wires were analyzed in the past with the help of Monte Carlo method [3,4]. Recently, shot noise for molecular devices has been studied from first principles (within density-functional formalism) [5] and also using parametric approach (within "orthodox" theory) [6].

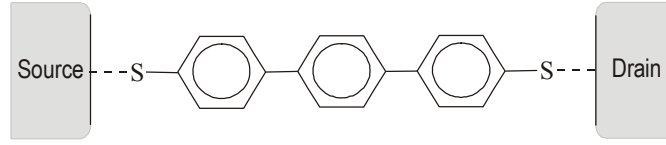

Fig.1 A schematic model of an ideal sample.

The main purpose of this work is to provide a systematic study of the current and its fluctuations of a particular class of conjugated molecules connecting a pair of metallic electrodes (see Fig.1). Among all the molecules, particularly important examples of molecular wires are organic polymers with extended π-conjugation (electron delocalization along the length of the entire molecule) [7]. Molecular wires consist of a linear chain of one or more parasubstituted benzene rings terminated by thiol linkages [8]. This end group ensures readily attachment to metal surfaces [9]. Since only delocalized π-electrons dictate the transport properties of analyzed structures, molecular system is described with the help of the single-orbital π-electron approximation, while the coupling to the electrodes is treated through the use of Newns-Anderson chemisorption theory [3,4,10-12].

## II. Methodology

The problem we are facing now is to solve a problem of electronic conduction between two continuum reservoirs of states (electrodes) via discrete energy levels of the molecule (bridge). Conceptually simple and transparent method is based on the scattering theory, where the current flowing through the device is depicted as a single electron scattering process [13]. Under the influence of bias voltage, the electron traveling from the source to drain is scattered by a molecule, which plays the role of a strong defect in perfectly periodic system of ideal electrodes. If the transmission probability $T(\varepsilon)$ for molecular system is known, the current can be calculated from the integration procedure [13]:

$$I = \frac{2e}{h} \int_{-\infty}^{+\infty} d\varepsilon T(\varepsilon)[f_S - f_D], \qquad (1)$$

where: $f_{S/D} = f(\varepsilon - \mu_{S/D})$ denotes the equilibrium Fermi distribution function with electrochemical potentials defined as $\mu_{S/D} = \varepsilon_F \pm eV/2$. In our non-self-consistent approach we must postulate potential distribution along the molecular bridge. The voltage is assumed to be dropped entirely at the molecule/electrode interfaces of the junction and remains constant (zero) in the molecule. Such assumption is justified by some self-consistent calculations [11], as proposed earlier to obtain a good fit to experimental data [14].

The general expression for the noise power of the current fluctuations in a two-terminal device is given by the expression [1]:

$$S = \frac{4e^2}{h} \int_{-\infty}^{+\infty} d\varepsilon \{T(\varepsilon)[f_S(1-f_S)+f_D(1-f_D)] + T(\varepsilon)[1-T(\varepsilon)](f_S-f_D)^2\}. \qquad (2)$$

Here the first two terms are the equilibrium noise contributions, and the third term (second order in the distribution function) is non-equilibrium or shot noise contribution to the power spectrum. In general, the theory of shot noise in nanoscopic systems allows us to define the so-called Fano factor with the help of relation [15-17]:

$$F = \frac{S}{2e\ I}. \qquad (3)$$

When $F = 1$, the magnitude of the shot noise reaches well-known Poisson limit (absence of correlations among the charge carriers), and when $F < 1$ – sub-Poisson value of shot noise is achieved (electron correlations reduce the level of fluctuations below the Poisson limit).

Now the transport problem is restricted to the evaluation of the transmission probability, which can be solved within the transfer matrix method [13]:

$$T(\varepsilon) = \mathrm{tr}[\varGamma_S G \varGamma_D G^+], \qquad (4)$$

where: $\varGamma_{S/D} = i[\varSigma_{S/D} - \varSigma^+_{S/D}]$ is broadening function related to the lifetime of an electron in a molecular state, $G = [\varepsilon\ J - H - \varSigma_S - \varSigma_D]^{-1}$ is the molecular Green's function ($J$ denotes identity matrix) and $\varSigma_{S/D}$ is the self-energy term of the source/drain electrode. The molecular system is described by the use of a simple Hückel Hamiltonian:

$$H = \sum_{k,\sigma} \varepsilon_{k,\sigma} c^+_{k,\sigma} c_{k,\sigma} + \sum_{k,\sigma} \left( c^+_{k,\sigma} t_{k,k+1} c_{k+1,\sigma} + \mathrm{h.c.} \right). \qquad (5)$$

In the above relation: $\varepsilon_k$ is local site energy, $t_{k,k+1}$ is the nearest-neighbor hopping integral, and $c^+_{k,\sigma}$ ($c_{k,\sigma}$) are the creation (annihilation) operators for electron with spin $\sigma$ on site $k$, respectively.

**III. Results**

Now we proceed to apply presented formalism to polymeric chains and report some characteristic features of the current and its fluctuations in two different regimes (linear and nonlinear, respectively). Calculations throughout this work are performed using the standard energy parameters (given in eV): $\varepsilon_k = 0$ (reference energy), $t_{k,k+1} = -2.5$, $\varepsilon_F = 0$. Zero value of the Fermi energy corresponds to the assumption that this level is located exactly in the middle of the HOMO-LUMO gap of the isolated molecule [18]. The whole system is assumed to be at room temperature ($293$ K), but anyway all the results are qualitatively the same for temperatures ranging from $0$ up to $350$ K.

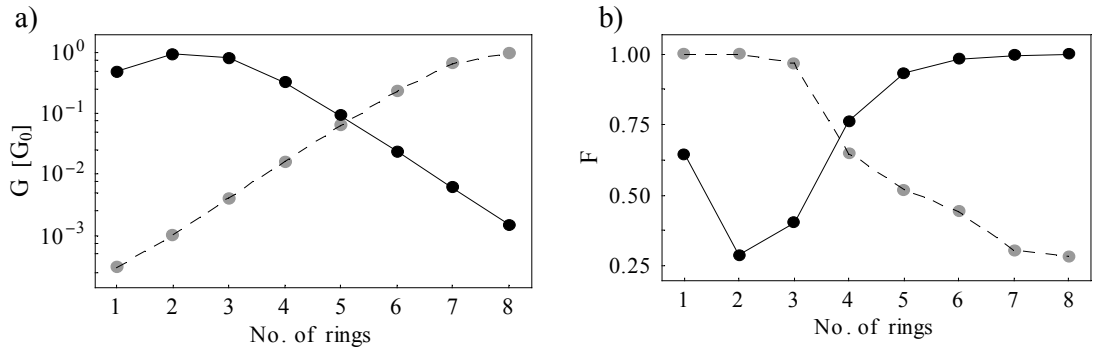

Fig.2 Dependence of the conductance (a) and Fano factor (b) on the number of benzene rings in the wire for the case of weak ($\Delta = 0.01$ eV – broken lines, grayish circles) and strong ($\Delta = 0.50$ eV – solid lines, black circles) coupling to the electrodes in the linear response regime ($V = 100$ mV).

Molecular wire is used to connect two ideal electrodes. The influence of the continua of reservoir states on the discrete energy levels of the molecule is incorporated in the self-energy terms ($\Sigma$). The effect of that coupling is to shift (real parts of $\Sigma$) and to broaden (imaginary parts of $\Sigma$) molecular energy levels. Since we restrict ourselves to qualitative effects of the molecular electronic structure and for the sake of simplicity, we have assumed that both electrodes are identical and their energy bands are half-filled [3,4,10-12]. Within this assumption, self-energy terms possess only its imaginary part, which is non-zero only for terminal (sulfur) atoms: $\Sigma_S = \Sigma_D = -i\Delta$. Here $\Delta = \beta^2/\gamma$ is the chemisorption coupling depending on the hopping parameter between the linking atom and the surface of the electrode ($\beta$) and the reservoir energy bandwidth ($4\gamma$). In our simplified approach, $\Delta$ is considered to be energy and voltage independent. However, our essential conclusions can be generalized well beyond this approximation, including treatment of the electrodes in more realistic description.

## A. Linear regime

At low bias voltages (small enough to treat the current as a linear function of an applied bias $I = GV$), the conductance $G$ of molecular junction is given through the Landauer formula [19]:

$$G = G_0 T(\varepsilon_F), \qquad (6)$$

where: $G_0 = 2e^2/h \approx 77.5$ [$\mu$S] is quantum of conductance and $T(\varepsilon_F)$ is the transmission function determined in the Fermi level $\varepsilon_F$ of the electrodes. Thus, the prediction of the conductance $G$ depends on the estimated position of the Fermi level relatively to the electronic spectrum, on exact conformation of the molecule (in this framework molecular wire is treated as a rigid structure) and on the strength of the coupling to the electrodes.

In Fig.2 we plot conductance G and the Fano factor F as a function of the number of benzene rings in the wire for an external bias $V = 100$ mV (linear response regime). In the limit of weak coupling we can observe Poissonian ($F = 1$) to sub-Poissonian ($F < 1$) crossover in the shot noise with increasing number of benzene rings in the polymeric chain. At the same time, the conductance increases almost exponentially. In the strong-coupling case, the behavior of discussed dependences is exactly opposite. The origin of such differences is due to the changes in the transmission function with increasing of number of benzene rings in the wire in the vicinity of the Fermi energy level (see Fig.3) [20].

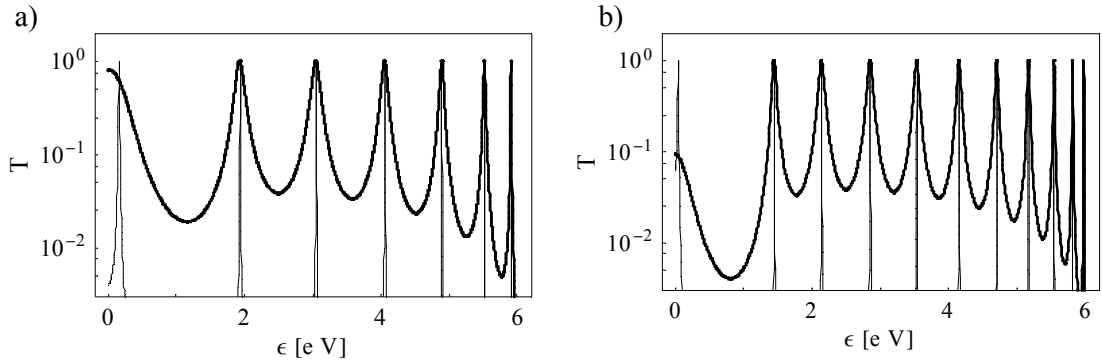

Fig.3 Transmission $T(\varepsilon) = T(-\varepsilon)$ as a function of electron energy for the case of three (a) and five (b) benzene rings in the limit of weak ($\Delta = 0.01$ eV – thin lines) and strong ($\Delta = 0.50$ eV – thick lines) coupling to the electrodes.

Here we also should not forget about the behaviour of transmission function during the band formation in a periodic structure including many benzene rings in series [21].

In addition, observed characteristics in Fig.2b inform us also about statistical properties of charge carriers (as mentioned in the previous section). Since we neglected all the electron-electron interactions within the molecular system (it is deficiency of the Hückel theory), such correlations are associated with the Pauli exclusion principle only [22]. Electrons travel independently from each other for: (i) short chains (one or two rings), while the coupling is weak or (ii) longer chains (more than six rings), while the coupling is relatively strong. In other extreme cases, electrons can not be treated as free particles anymore.

### B. Nonlinear regime

At higher bias voltages ($V > 100$ mV), the current is nonlinear function of the voltage, because of the exponential dependence of the Fermi distribution functions on applied bias (transmission $T(\varepsilon)$ is assumed to be bias-independent). A sequence of irregularly placed steps in the I-V characteristics is attributed to the discreteness of the molecular levels [23]. Since electrons can travel from one electrode (source) to another (drain) using molecular orbitals only, a sharp increase in the current occurs whenever chemical potential of the metal aligns with the molecular energy level.

However, one can indicate the general tendency: the stronger molecule-to-electrodes coupling, the larger the current flowing through the device, and the smoother I-V curve [20]. The shape and the height of the current steps are affected by the broadening effects in the electronic structure of the molecule due to its coupling with the electrodes. In particular, the height of that step is directly proportional to the area of the corresponding peak in the transmission spectrum. In the weak-coupling case ($\beta \ll t_{k,k+1}$), transmission peaks are very narrow and their location coincide with molecular resonances (see Fig.3). Therefore, after integration we obtain a step-like dependence of the I-V curve and the corresponding current is relatively small. In the strong-coupling limit ($\beta \sim t_{k,k+1}$), transmission peaks gain a substantial width and may even merge for the states close in energy (see Fig.3). As a consequence, the current is a fairly smooth function of bias voltage and the values of the current are higher. Such relations explain the differences in the magnitude of the current flowing through the device and simultaneously its fluctuations in two border regimes.

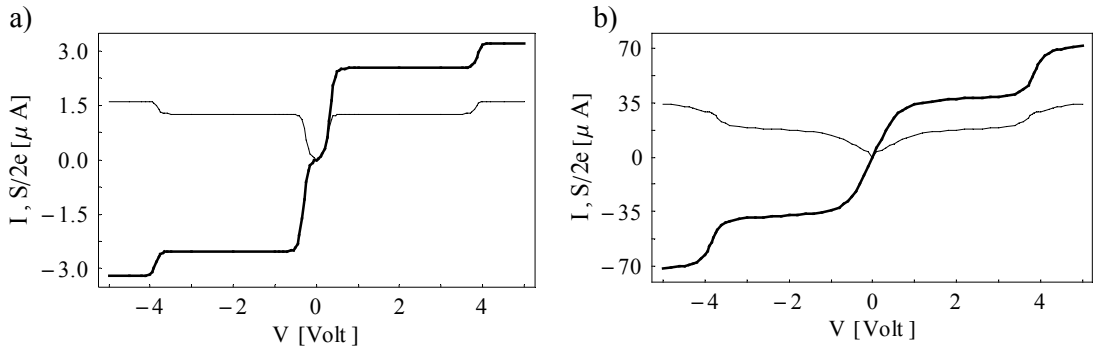

Fig.4 I-V characteristics (thick lines) and noise power (thin lines) for three benzene rings in the chain for the case of weak (a: $\Delta = 0.01$ eV) and strong (b: $\Delta = 0.50$ eV) coupling to the electrodes in nonlinear response regime.

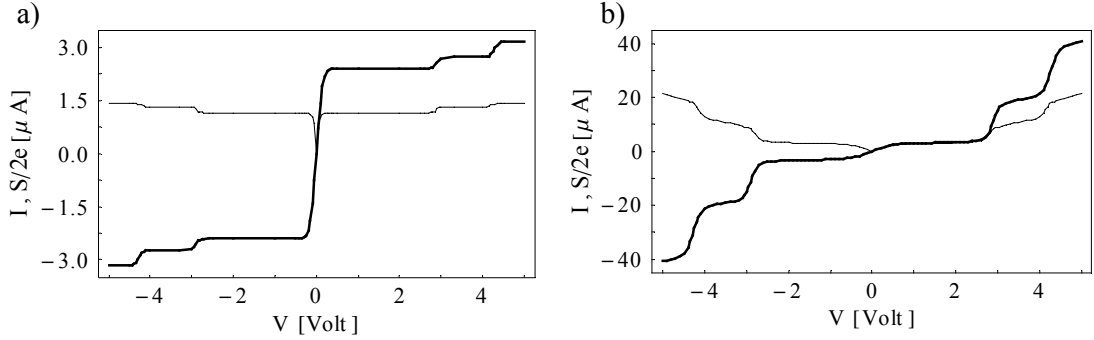

Fig.5 The same as Fig.4, but for five benzene rings in the wire.

Figures 4 and 5 show the dependence of I-V characteristics and current fluctuations on the coupling strength. In the limit of the weak coupling, the transition in the shot noise from Poissonian to sub-Poissonian is always observed after the first step in the I-V dependence (see Fig.4a and 5a). It means that electrons are correlated (one electron feels the presence of the others owing to Pauli exclusion principle) after the tunnelling process has occurred. Moreover, the current is twice as large as noise power in the sub-Poissonian area, independently of the number of benzene rings in the wire (where noise-to-current ratio $S/I$ is approximately equal to $1/2$). From such observation we can conclude that in the weak-coupling regime and sub-Poissonian area, the lowest possible noise remains on the same level independently from the length of polymeric chain.

In the strong-coupling case we observe decrease of the current and noise power with increasing number of benzene rings in the wire (compare Fig.4b and 5b). Another striking feature is associated with Poissonian to sub-Poissonian crossover in the noise power spectrum. There is no such transition in the case of short polymeric wires (see Fig.4b), since current fluctuations are already in the sub-Poissonian area, where electron correlations are important. However, results for longer chains still reveal such crossover after the first step in the I-V characteristic (see Fig.5b). Generally, current fluctuations are less significant for longer polymeric chains then for shorter ones in the strong-coupling regime.

## IV. Summary

In this work we have studied I-V characteristics and their fluctuations for a wide class of polymeric chains inserted between two electrodes in two different regimes: linear and nonlinear, respectively. The results of our calculations indicate strong dependence of the shot noise on the molecule-to-electrodes coupling. According to our predictions, in the linear response regime we can observe Poissonian to sub-Poissonian crossover in the shot noise with increasing number of benzene rings in the polymeric chain (weak-coupling case) or exactly opposite transition (strong-coupling case). However, in nonlinear response regime, Poissonian to sub-Poissonian crossover in the shot noise as a function of applied voltage can be present or absent. The final result is determined by the combined effect of the length of polymeric chain and its strength of the coupling with the electrodes.

Presented method is based on few drastic approximations and obtained results should be considered as qualitative only. Namely, we have assumed that the voltage drops take place entirely at the metal/molecule contacts. However, in the real systems the electrostatic potential spatial profile may be much more intricate. Particularly, at higher voltages the electronic structure of the molecule can be changed due to the electric field (Stark effect). Within our calculational scheme, it is possible to include this phenomenon by assuming phenomenological forms of the potential drop inside the molecular bridge [23-25]. Moreover, molecular system is described within simple Hückel model, where all the effects of electron-electron interactions are completely neglected. When the bias is applied, the charge redistribution inside the molecule and screening effects near the electrodes could have a significant influence on transport properties of the molecular devices. In this case, better results are expected in the self-consistent procedure, in which the charge density in the molecule and the positions of molecular energy levels are achieved [25, 26]. This aim can be realized by a direct generalization of our simple model.

It should be also mentioned that discussed formalism is valid only in the coherent transport regime, where the tunnelling electron is undisturbed by phase destroying thermal motions of the nuclei. Furthermore, for longer polymeric chains, the residence time of an injected electron on a molecular wire can be comparable with nuclear vibrations ($\sim 10^{-12}$ s) and relaxation processes become important. In the presence of the dephasing processes a more general expression for the current should be used [27]. Furthermore, increasing the number of benzene rings we are closer and closer to the so-called diffusive transport regime.

Although such simplifications, we believe that the algorithm presented here for the noise power calculations is efficient theoretical tool in designing molecules with the lowest possible noise, especially for short polymers at low voltages. Anyway, further investigations are expected to confirm our predictions in more realistic description of the whole electrode/molecule/electrode system.

## Acknowledgements

Author is very grateful to B. Bułka, T. Kostyrko and B. Tobijaszewska for illuminating discussions. Special thanks are also addressed to S. Robaszkiewicz for his stimulating suggestions.